\documentclass[preprint,preprintnumbers,amsmath,amssymb]{revtex4}
\usepackage[abs]{overpic}
\usepackage{graphicx,subfigure}
\preprint{HUPD1405}
\begin{document}
\def\tbr{\textcolor{red}}
\def\tcr{\textcolor{red}}
\def\ov{\overline}
\def\dprime{{\prime \prime}}
\def\nn{\nonumber}
\def\f{\frac}
\def\p{\partial}
\def\H{\mathcal{H}}
\def\beq{\begin{equation}}
\def\eeq{\end{equation}}
\def\bea{\begin{eqnarray}}
\def\eea{\end{eqnarray}}
\def\bsub{\begin{subequations}}
\def\esub{\end{subequations}}
\def\dc{\stackrel{\leftrightarrow}{\partial}}
\def\d{\partial}
\def\sla#1{\rlap/#1}
\def\mH{\mathscr{H}}
\def\tD{\tilde{D}}
\def\Q{{\cal Q}}
\def\mpim{m_{\pi^-}}
\def\mpi0{m_{\pi^0}}
\def\meta{m_\eta}
\def\tauepp{\tau^- \to \nu_\tau \eta \pi^- \pi^0}
\def\Vcurrent{\bar{d}\gamma_\mu u}
\def\Acurrent{\bar{d} \gamma_\mu \gamma_5 u}
\def\VmA{\bar{d}\gamma_\mu(1-\gamma_5)u}
\def\epp{\eta \pi^- \pi^0}
\def\hg{{G}_{2p}}
\def\nn{\nonumber}
\def\beq{\begin{equation}}
\def\eeq{\end{equation}}
\def\bei{\begin{itemize}}
\def\eei{\end{itemize}}
\def\bea{\begin{eqnarray}}
\def\eea{\end{eqnarray}}
\def\ynu{y_{\nu}}
\def\ydu{y_{\triangle}}
\def\ynut{{y_{\nu}}^T}
\def\ynuv{y_{\nu}\frac{v}{\sqrt{2}}}
\def\ynuvt{{\ynut}\frac{v}{\sqrt{2}}}
\def\s{\partial \hspace{-.47em}/}
\def\ad{\overleftrightarrow{\partial}}
\title{Study of an anomalous tau lepton decay 
using \\ a chiral Lagrangian with vector mesons
}
\author{Daiji Kimura}
\address{Ube National College of Technology, Ube  Yamaguchi 755-8555, Japan}
\author{Takuya Morozumi, Hiroyuki Umeeda}
\address{Graduate School of Science, Hiroshima University  Higashi-Hiroshima
739-8526, Japan}
\begin{abstract}
The hadronic tau decay  $\tau^- \to \nu_\tau \eta \pi^- \pi^0$  occurs
through V-A weak current. In this decay mode, the vector current contribution is intrinsic parity violating and the axial current contribution is G parity violating. The latter contribution is suppressed due to tiny isospin breaking.  We have computed  both  vector and axial vector form factors
using a chiral Lagrangian with vector mesons including the effect of isospin breaking and intrinsic parity violation. A numerical result of 
the invariant mass distribution is shown and the structure of $\rho$ 
resonance can be seen in the distribution with respect to $M_{\pi^- \pi^0}$.  
\end{abstract}
\maketitle
\vspace*{5pt}
\section{Introduction: Intrinsic Parity violation of $\tau$ decay}
Many hadronic $\tau$ decay modes are found.
Among them, 
decay modes with more than two pseudo-Nambu Goldstone bosons as a final state  
are interesting since they include both intrinsic 
parity (IP) violating and conserving amplitudes \cite{Li:1997tj},\cite{Kuhn:1992nz},\cite{Decker:1992rj},\cite{Dumm:2012vb}.
IP for bound state of quark and anti-quark is assigned to 
scalar, pseudo-scalar, vector and axial vector mesons.
\bea
&{\rm IP}=+1,&S_{ij}=\bar{q_j} q_i, V_{\mu ij}=\bar{q_j} \gamma_\mu q_i,  \\
&{\rm IP}=-1,&P_{ij}=i \bar{q_j} \gamma_5 q_i,
A_{\mu ij}=\bar{q_j} \gamma_\mu \gamma_5 q_i.\nn 
\eea 
With this assignment, IP is the sign which one obtains 
by replacing $\gamma_5$ with 
$-\gamma_5$.  As for gauge bosons, which are not bound state of quark and anti-quark, vector gauge boson is regarded as,
\bea
V_\mu=\frac{A_{L \mu}+A_{R \mu}}{2},
\label{eq:photon}
\eea 
\\
\\
where $A_L$ ($A_R$) denotes the gauge boson 
which couples to left(right)-handed quark. 
The IP is defined as the sign which one obtains by interchanging
 $A_L$ and $A_R$ in Eq.(\ref{eq:photon}). 
Therefore, IP of a single photon is $+1$.
\begin{table}[htbp]
\begin{center}
\begin{tabular}{|c|c|c|c|c|} \hline
$$      & G    &  $e^{i \pi I_y}$ & $C$ & IP\\ \hline
$\pi^+$ & $-1$ & $\pi^-$ & $-\pi^-$ & $-1$\\
$\pi^0$ & $-1$ & $-\pi^0$ & $\pi^0$ &$-1$\\ 
$\pi^-$ & $-1$ & $\pi^+$ & $-\pi^+$ &$-1$\\
$\eta_8$ & $+1$ & $\eta_8$ & $\eta_8$ &$-1$\\
$\bar{d} \gamma_\mu u$ &  $ +1$ & 
 $-\bar{u} \gamma_\mu d$ &
$-\bar{u} \gamma_\mu d$ & $+1$ \\ 
$\bar{d} \gamma_\mu \gamma_5  u$ &  $-1$ &  $-\bar{u} \gamma_\mu \gamma_5 d$ & 
$\bar{u} \gamma_\mu \gamma_5 d$ &$-1$\\ \hline
\end{tabular}
\caption{Assignment of G parity and Intrinsic Parity (IP) on $\pi$,$\eta_8$
,vector and axial vector current}
\end{center} 
\label{tab1}
\end{table}
\label{}
The hadronic current of the weak decay amplitude for $\tauepp$ is
 written in terms of 
the sum of the matrix element of vector
current and the matrix element of axial current.
The IP of the (axial) vector current is $+1(-1)$ and 
the IP of the final state;
$\eta \pi^0 \pi^-$ is $-1$. 
The vector current contribution corresponds to
the intrinsic parity violating process and the axial current contribution is
intrinsic parity conserving one.   
 
Next we consider another parity called 
G parity. G parity is defined as $C e^{i \pi I_y}$.
As shown in Table \ref{tab1}, G parity for $\eta$ is $+1$ which is different 
from pion's G parity.  
 G parity and IP of vector current are the same to each other 
and is $+1$.
Their assignment on axial current is also the same and it is  $-1$.
One also notes that pion's G parity and IP are the same to each other 
and they are $-1$.
Therefore vector current contribution is G parity conserving and IP violating while the axial current 
contribution is G parity violating and IP conserving. 
\section{Interaction and Form factors}
The weak interaction which is relevant for $\tauepp$ is given as,
\bea
-{\cal L}_{\rm int}=\frac{G_F}{\sqrt{2}}V_{ud}^\ast \overline{\nu}_\tau
\gamma_\mu (1-\gamma_5) \tau \bar{d} \gamma_\mu (1-\gamma_5) u. \nn \\ 
\eea
The hadronic matrix elements for vector and axial vector currents
between vacuum and $\eta \pi^- \pi^0$ are written as,
\bea
\langle \epp|\Vcurrent|0 \rangle &=& 
-i V_{3 \mu} F_3
, \nn \\ 
\langle \epp|\Acurrent|0 \rangle &=& V_{1\mu} F_{1}\nn \\
&+&
V_{2 \mu} F_{2}+V_{4 \mu} F_{4},
\eea
where $F_i$ ($i=1-4$) are form factors and  $V_{i \mu}$  are 
written in terms of the four vectors of mesons in the final states.
\bea
V_{1 \mu}&=& (p_{\pi^-}- p_{\eta})_\mu-Q_\mu 
\frac{(p_{\pi^-}- p_{\eta}) \cdot Q}{Q^2}, \nn \\ 
V_{2 \mu}&=& ( p_{\pi^0}-p_\eta)_\mu-Q_\mu \frac{(p_{\pi^0}- p_{\eta}) 
\cdot Q}{Q^2}, \nn \\ 
{V_{3\mu}} &=& {\epsilon_{\mu \nu  \rho  \sigma} p^{\nu}_{\pi^-} p^\rho_{\pi^0} p^\sigma_{\eta}},  \nn \\
V_{4 \mu}&=&Q_\mu \equiv (p_{\pi^0}+p_\eta+p_{\pi^-})_\mu,
\eea
where $\epsilon^{0123}=1$. The form factor $F_3$ for vector current is G parity conserving and 
IP violating. The axial vector form factors $F_1, F_2$, and $F_4$
are G parity violating and IP conserving.
\section{Isospin breaking and $\pi_0$ and $\eta (\eta^\prime)$ mixing}  
To compute the matrix element for the hadronic current, 
we introduce a chiral Lagrangian with vector mesons.
The axial vector form factor becomes non-zero due to the isospin violation
and we must keep the mass difference of up quark and down quark.
Since the IP is conserved 
for the matrix element for the axial current, we adopt the IP conserving 
part of Lagrangian \cite{Kimura:2014wsa},
\bea
{\cal L}&=&\frac{f^2}{4} {\rm Tr} (D_\mu U D^\mu U^{\dagger}) \nn \\
&+&B {\rm Tr}[M_q (U+U^{\dagger})] \nn \\
&-&ig_{2p}{\rm Tr}(\xi M_q \xi 
-\xi^\dagger M_q \xi^\dagger)\cdot \eta_0 \nn \\
&+&\frac{1}{2} 
\partial_\mu \eta_0 \partial^\mu \eta_0-\frac{M_0^2}{2}\eta_0^2 \nn \\
&+& M_{V}^2 {\rm Tr}\left[\left(V_{\mu}-\frac{\alpha_{\mu}}{g}
\right)^2 \right],
\label{eq:lag} 
\eea
where we use SU(3) octet mesons which are given by $3 \times 3$ matrix;
$\pi=\sum_{a=1}^8 \pi^a T^a$ to write the pseudo Nambu Goldstone bosons.
The fields which appear in Eq.(\ref{eq:lag}) are given as, 
\bea
\xi&=&\exp(i \frac{\pi}{f}), \quad U=\xi^2, \nn \\
D_{\mu}U&=&(\p_\mu+iA_{L\mu})U -i U A_{R \mu}, \nn \\
D_{\mu} U^\dagger&=&(\p_\mu + i A_{R \mu}) U^\dagger -i U^\dagger 
A_{L \mu}, \nn \\
\alpha_{\mu}&=&\frac{1}{2i}(\xi^\dagger D_{L\mu}\xi+\xi D_{R\mu} \xi^\dagger),
\eea
where $D_{L\mu}=\p_\mu  +i A_{L \mu}$ and $D_{R\mu}=\p_\mu  +i A_{R \mu}$.
$V_\mu$ denotes SU(3) vector mesons octet including $\rho, K^\ast$, and 
$\omega_8$. $\eta_0$ denotes SU(3) singlet pseudo-scalar meson.
The term proportional to $g_{2p}$ leads to octet and singlet mixing such as
$\pi_0-\eta_0$ mixing and $\eta_8-\eta_0$ mixing. 
The Lagrangian Eq.(\ref{eq:lag}) is invariant under replacement $\pi \to -\pi,
(A_L, A_R) \to (A_R, A_L)$ and it is IP conserving. 
We introduce 
isospin breaking by keeping the mass difference of up quark and down quark
and the current quark mass $M_q$ is given as, 
\bea
M_q=\begin{pmatrix} m_u & 0 & 0 \\
                  0 & m_d & 0 \\
                  0 & 0 & m_s \end{pmatrix}.
\eea
$\pi_0$ and the other isosinglet mesons $\eta, \eta^\prime$ are mixed. 
Since the axial vector form factor is  G parity violating and very sensitive to the mixing matrix of the neutral mesons, we investigate their mass squared matrix,
\bea
M^2=
\begin{pmatrix}M^2_{\pi^+} & \frac{\Delta_K}{\sqrt{3}} &  -
{G}_{2p}\Delta_K \\ 
\frac{\Delta_K}{\sqrt{3}} & 
\frac{2\Sigma_K-M^2_{\pi^+}}{3}  &  \frac{G_{2p}(\Sigma_K-
2M^2_{\pi^+})}{\sqrt{3}} \\
-
{G}_{2p}\Delta_K&  \frac{{G}_{2p}(\Sigma_K-
2M^2_{\pi^+})}{\sqrt{3}}  & M_{0}^2
\end{pmatrix},&& \nn \\ 
\label{Eq:M2}
\eea
where  $\Delta_K=M^2_{K^+}-M^2_{K^0}, \Sigma_K=M^2_{K^+}+M^2_{K^0}$
and $\hg=\frac{g_{2p}f}{B}$.  We note $\Delta_K$ denotes the isospin breaking 
and is proportional to $m_u-m_d$.
The mass matrix has two parameters $\hg$ and $M_{0}^2$.
The latter is constrained by the relation,
\bea
{\rm Tr}(M^2)= M_{\pi^0}^2+M_{\eta}^2+M_{\eta^\prime}^2. 
\eea
 We diagonalize Eq.(\ref{Eq:M2}) and determine the mixing matrix. 
Using the masses of pseudo-scalar mesons $M_{K^+}, M_{K^0}, M_{\pi^+}, M_{\pi^0}, M_{\eta}$ and $M_{\eta^\prime}$, one obtains 
the mass eigenvalues as function of the dimensionless parameter $\hg$. In Fig.1,
we show $\Delta m=$ mass eigenvalue - physical mass for $\pi^0$, $\eta$ and $\eta^\prime$
 as a function $\hg$.  We find that one can reproduce masses of $\eta$ and $\eta^\prime$ for $|\hg|\sim 0.43$. However,
the predicted neutral pion mass is higher than its physical value and the discrepancy is about 5 (MeV). We expect the 
discrepancy could be remedied if we have introduced the electromagnetic effect and leave it for future study. Ignoring the
discrepancy, one obtains
the mixing matrix $O$ which relates original basis to  the mass eigen-states. 
\bea
\begin{pmatrix} \pi_0 \\ \eta_8 \\ \eta_0
\end{pmatrix}=\begin{pmatrix} O_{11} & O_{12} &O_{13} \\
O_{21} & O_{22} &O_{23} \\ O_{31} & O_{32} &O_{33} 
\end{pmatrix}
\begin{pmatrix}\pi^0 \\ \eta \\
\eta^\prime
\end{pmatrix}, 
\eea
\bea
O=
\begin{pmatrix} 0.99996 & -0.00900 & -0.00137 \\
  0.00859& 0.98263 & -0.18538 \\
  0.00302 & 0.18536 & 0.98267 \end{pmatrix},&&
\eea
where we have shown elements of $O$ for $\hg=-0.43$.
Having determined the mixing matrix, one can compute Feynman diagrams
which contribute to the form factors for axial vector current; $F_1, F_2$ and $F_4$.
The amplitude is proportional to off-diagonal elements
of $O$ and suppressed by tiny isospin violation
and vanishes in the isospin limit. 
\begin{figure}
\begin{center}
\includegraphics[height=6.7cm,width=7.0cm,clip]{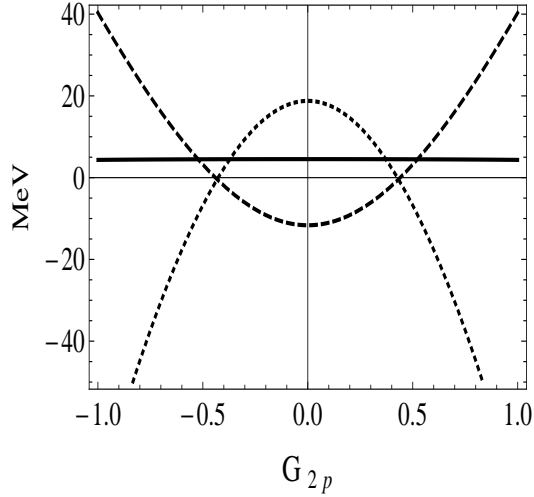}
\caption{$\Delta m$=eigenvalue$-$physical value for neutral meson masses.
The dashed line is  $\Delta m$ for $\eta^\prime$, the dotted line is for $\eta$
and the solid line is for $\pi_0$.
The horizontal axis is the coupling constant $G_{2p}$.} 
\end{center}
\end{figure}
\section{Vector form factor and intrinsic parity violation}
The intrinsic parity violating interaction contributes to 
the vector form factor. There are two categories of interactions
with intrinsic parity violating effect. One is Wess Zumino term,
\cite{Wess:1971yu} \cite{Pich:1987qq}
\bea
&&\mathcal{L}_{WZ}=\frac{i}{\pi^2 f^3}\epsilon^{\mu\nu\rho\sigma}
\mathrm{tr}V^{\rm ext}_\mu \partial_{\nu}{\pi}
\partial_{\rho}{\pi}\partial_{\sigma}{\pi},
\eea
and it contributes to the vector form factor $F_3$
\bea
\langle \pi^- \pi^0 \eta|\bar{d}\gamma^\mu u|0 \rangle
&=&\frac{(O_{11}O_{22}-O_{12}O_{21})}{2\sqrt{6}\pi^2f^3} \nn \\
&\times& \epsilon^{\mu\nu\rho\sigma}p^-_\nu p^0_\rho 
p^\eta_\sigma,
\eea
as,
\bea
F_{3 WZ}=i\frac{(O_{11}O_{22}-O_{12}O_{21})}{2\sqrt{6}\pi^2f^3}.
\eea
The other intrinsic parity violating part including the vector meson
contribution is obtained from \cite{Fujiwara:1984mp,Bando:1987br}.
Some of the terms in \cite{Fujiwara:1984mp} are not
charge conjugation invariant and we use the following terms,    
\bea
\mathcal{L}_{IPV}&=&\sum_{i=1,2,4} C_i \mathcal{L}_i,\\
\mathcal{L}_1&=&i\epsilon^{\mu\nu\rho\sigma}\mathrm{Tr}[\alpha_{L\mu}
\alpha_{L\nu}\alpha_{L\rho}\alpha_{R\sigma}-(R\leftrightarrow L)],
\label{mat14}\nn \\
\mathcal{L}_2&=&i\epsilon^{\mu\nu\rho\sigma}\mathrm{Tr}[\alpha_{L\mu}
\alpha_{R\nu}\alpha_{L\rho}\alpha_{R\sigma}],\label{mat15} \nn \\
\mathcal{L}_4&=&-\frac{1}
{2}\epsilon^{\mu\nu\rho\sigma}\mathrm{Tr}F_{V\mu\nu}
[\alpha_{L\rho}\alpha_{R\sigma}-(R\leftrightarrow L)],\label{mat16} \nn
\eea
where
\bea
&&\alpha_{L\mu}=\alpha_\mu+\alpha_{\perp\mu}-gV_\mu,  \nn \\
&&\alpha_{R\mu}=\alpha_\mu-\alpha_{\perp\mu}-gV_\mu, \nn \\
&&\alpha_{\mu}=\frac{1}{2i}(\xi^\dagger D_{L\mu}\xi+\xi D_{R\mu}\xi^\dagger)\label{def3},\nn \\
&&\alpha_{\perp\mu}=\frac{1}{2i}(\xi^\dagger D_{L\mu}\xi-\xi D_{R\mu}\xi^\dagger)\label{def4}.
\eea
See \cite{Dumm:2012vb} with another resonance chiral Lagrangian.
\section{The hadronic invariant mass distribution}
Following \cite{Kuhn:1992nz}, the hadronic invariant mass distribution is given
as,
\bea
&& \frac{d^3 Br}{d M_{\pi^- \pi^0} d M_{\pi^- \eta} d M_{\pi^- \pi^0 \eta}}
=\frac{G_F^2 |V_{ud}|^2}{(2 \pi)^5}\frac{M_{\pi^- \pi^0} M_{\pi^- \eta}}{M_{\pi^- \pi^0 \eta}} \nn \\
&& \frac{(m_\tau^2-M^2_{\pi^- \pi^0 \eta})^2}{
16 m_\tau^3 \Gamma_{\tau}}
\times  
\sum_{X=A,B,SA} <\bar{L}_{X}> W_X,\label{eq:dBr}
\eea
\bea
\sum_{X=A,B,SA}<\bar{L}_{X}> W_X&=&\left(\frac{2}{3}+\frac{m^2_\tau}{3 Q^2}
\right)(W_A+W_B) \nn \\&+&\frac{m^2_\tau}{Q^2} W_{SA}, \quad
\eea
where $Q^2=M^2_{\pi^- \pi^0 \eta}$ and  $W_X (X=A,B,SA)$ are defined as,
\bea
W_A&=&|x_1 F_1+x_2 F_2|^2\nn \\
    &+&x_3^2 \Bigl{[}|F_1-\frac{F_2}{2}|^2+\frac{3 |F_2|^2}{4} \Bigr{]},\nn \\
W_B&=&x_4^2 |F_3|^2,\quad W_{SA}= Q^2 |F_4|^2,
\eea 
where the kinematical variables $x_1 \sim x_4$ are defined through
the energies and momenta of hadrons in the hadronic rest frame.
\bea
E_\eta&=&\frac{Q^2-M^2_{\pi^- \pi^0}+m_\eta^2}{2 \sqrt{Q^2}},
q_{\eta x}= \sqrt{E_\eta^2-m_\eta^2}, \nn \\
E_{\pi^0}&=&\frac{Q^2-M^2_{\pi^- \eta}+m_{\pi^0}^2}{2 \sqrt{Q^2}}, \nn \\
q_{\pi^0 x}&=&\frac{2 E_{\pi^0} E_\eta -M^2_{\pi^0 \eta}+m^2_{\pi^0}+m^2_{\eta}}{2 q_{\eta x}},
\nn \\
E_{\pi^-}&=& \sqrt{Q^2}-E_{\pi^0}-E_\eta, \nn \\ 
q_{\pi^- x}&=&\frac{2 E_{\pi^-} E_\eta-M^2_{\pi^- \eta}+ m^2_{\pi^-}+m^2_{\eta}}{2 q_{\eta x}}, \nn \\
x_1&=& q_{\pi^- x}-q_{\eta x},
x_2= q_{\pi^0 x}-q_{\eta x}, \nn \\
x_3&=& q_{\pi^- y}=-q_{\pi^0 y}=-\sqrt{E^2_{\pi^0}-q^2_{\pi^0 x}-m^2_{\pi^0}},\nn \\
x_4&=&\sqrt{Q^2} x_3(-q_{\pi^0 x}-q_{\pi^-x}).
\eea
In Fig.2, we have plotted the differential branching fraction as a function of $M_{\pi^- \pi^0}$ for fixed $M_{\pi^- \pi^0  \eta}$ and $M_{\pi^- \eta }$.
We set $C_1=C_2=C_4=1$. 
\begin{figure}
\begin{center}
\includegraphics[height=6.0cm,width=7.0cm,clip]{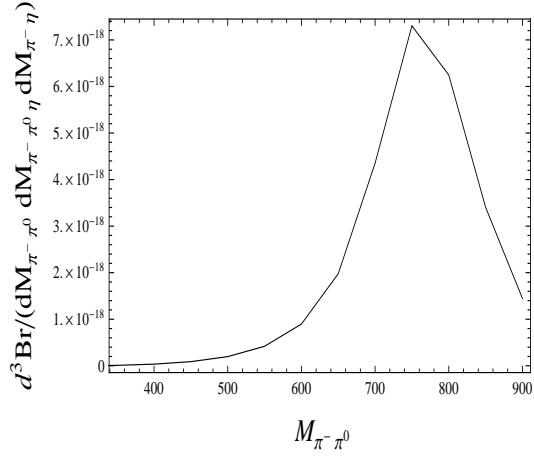}
\caption{Hadronic invariant mass distribution
Eq.(\ref{eq:dBr}).
The unit is MeV$^{-3}$ and invariant masses are 
$300$ (MeV)$< M_{\pi^- \pi^0} < 900$ (MeV),
$M_{\pi^- \pi^0 \eta}=1500$ (MeV) and  $M_{\pi^- \eta}=900$ (MeV).} 
\end{center}
\end{figure}
\section{Conclusion}
\begin{itemize}
\item[1] We have studied intrinsic parity violating process 
$\tau^- \to \eta \pi^- \pi^0 \nu$ taking G parity violating 
effect into account.
\item[2] A chiral Lagrangian with explicit isospin breaking and intrinsic parity violation is adopted.  With this Lagrangian, 
the non-vanishing contribution to the
axial vector form factor ($F_1,F_2,F_4$) are computed. The intrinsic
parity violating interactions are included and the vector form factor $F_3$ is computed. The IP violating interaction beyond Wess Zumino term includes 
three coefficients $C_1,C_2, C_4$.
\item[3]  As a preliminary numerical result, we showed 
the differential rate with respect to hadronic invariant mass for the case
$C_i=1 (i=1,2,4)$ and found the peak of $\rho^-$
meson in the distribution with respect to
$M_{\pi^- \pi^0}$. More serious studies which include the comparison with data of Belle \cite{Inami:2008ar} and the fitting of the coefficients $C_i$ using 
the data will be shown.
\end{itemize}








\end{document}